\documentclass[12 pt, amsfonts, amssymb,color]{article}

\usepackage{amsmath,amssymb,amsfonts,latexsym,float,graphics,epsfig}

\begin{document}

\renewcommand\theequation{\arabic{section}.\arabic{equation}}
\catcode`@=11 \@addtoreset{equation}{section}
\newtheorem{axiom}{Definition}[section]
\newtheorem{theorem}{Theorem}[section]
\newtheorem{axiom2}{Example}[section]
\newtheorem{lem}{Lemma}[section]
\newtheorem{prop}{Proposition}[section]
\newtheorem{cor}{Corollary}[section]
\newcommand{\be}{\begin{equation}}
\newcommand{\ee}{\end{equation}}

\newcommand{\equal}{\!\!\!&=&\!\!\!}
\newcommand{\rd}{\partial}
\newcommand{\g}{\hat {\cal G}}
\newcommand{\bo}{\bigodot}
\newcommand{\res}{\mathop{\mbox{\rm res}}}
\newcommand{\diag}{\mathop{\mbox{\rm diag}}}
\newcommand{\Tr}{\mathop{\mbox{\rm Tr}}}
\newcommand{\const}{\mbox{\rm const.}\;}
\newcommand{\cA}{{\cal A}}
\newcommand{\bA}{{\bf A}}
\newcommand{\Abar}{{\bar{A}}}
\newcommand{\cAbar}{{\bar{\cA}}}
\newcommand{\bAbar}{{\bar{\bA}}}
\newcommand{\cB}{{\cal B}}
\newcommand{\bB}{{\bf B}}
\newcommand{\Bbar}{{\bar{B}}}
\newcommand{\cBbar}{{\bar{\cB}}}
\newcommand{\bBbar}{{\bar{\bB}}}
\newcommand{\bC}{{\bf C}}
\newcommand{\cbar}{{\bar{c}}}
\newcommand{\Cbar}{{\bar{C}}}
\newcommand{\Hbar}{{\bar{H}}}
\newcommand{\cL}{{\cal L}}
\newcommand{\bL}{{\bf L}}
\newcommand{\Lbar}{{\bar{L}}}
\newcommand{\cLbar}{{\bar{\cL}}}
\newcommand{\bLbar}{{\bar{\bL}}}
\newcommand{\cM}{{\cal M}}
\newcommand{\bM}{{\bf M}}
\newcommand{\Mbar}{{\bar{M}}}
\newcommand{\cMbar}{{\bar{\cM}}}
\newcommand{\bMbar}{{\bar{\bM}}}
\newcommand{\cP}{{\cal P}}
\newcommand{\cQ}{{\cal Q}}
\newcommand{\bU}{{\bf U}}
\newcommand{\bR}{{\bf R}}
\newcommand{\cW}{{\cal W}}
\newcommand{\bW}{{\bf W}}
\newcommand{\bZ}{{\bf Z}}
\newcommand{\Wbar}{{\bar{W}}}
\newcommand{\Xbar}{{\bar{X}}}
\newcommand{\cWbar}{{\bar{\cW}}}
\newcommand{\bWbar}{{\bar{\bW}}}
\newcommand{\abar}{{\bar{a}}}
\newcommand{\nbar}{{\bar{n}}}
\newcommand{\pbar}{{\bar{p}}}
\newcommand{\tbar}{{\bar{t}}}
\newcommand{\ubar}{{\bar{u}}}
\newcommand{\utilde}{\tilde{u}}
\newcommand{\vbar}{{\bar{v}}}
\newcommand{\wbar}{{\bar{w}}}
\newcommand{\phibar}{{\bar{\phi}}}
\newcommand{\Psibar}{{\bar{\Psi}}}
\newcommand{\bLambda}{{\bf \Lambda}}
\newcommand{\bDelta}{{\bf \Delta}}
\newcommand{\p}{\partial}
\newcommand{\om}{{\Omega \cal G}}
\newcommand{\ID}{{\mathbb{D}}}
\newcommand{\pr}{{\prime}}
\newcommand{\prr}{{\prime\prime}}
\newcommand{\prrr}{{\prime\prime\prime}}

\newcommand{\bel}{\begin{equation}\label}

\title{ Branched Hamiltonians and time translation symmetry breaking in  equations of the Li\'{e}nard type}
\author{A Ghose-Choudhury\footnote{E-mail aghosechoudhury@gmail.com}\\
Department of Physics, Diamond Harbour Women's University,\\ D. H Road, Sarisha,
West-Bengal 743368, India\\
\and
Partha Guha\footnote{E-mail: partha@bose.res.in}\\
SN Bose National Centre for Basic Sciences \\
JD Block, Sector III, Salt Lake \\ Kolkata 700098,  India \\
}

\date{ }

 \maketitle

\smallskip

\smallskip

\begin{abstract}
\textit{ Shapere and Wilczek ( Phys. Rev. Lett. 109, 160402 and 200402 (2012)) have
recently described certain singular Lagrangian systems which display
spontaneous breaking of time translation symmetry. We begin by considering
the standard
Li´enard equation for which a Lagrangian is constructed by using the method
of Jacobi Last Multiplier. The velocity dependance of the Lagrangian is
such that the momentum may exhibit multivaluedness thereby leading to the
so called branched Hamiltonian. Next with a quadratic velocity dependance
in the Li\'enard equation one can construct a Hamiltonian description
involving a position dependent mass. We compute the Lagrangian and
Hamiltonian of this system and show that the canonical Hamiltonian is
single valued . However, we find that up to a constant shift, the square of
this Hamiltonian describes systems giving rise to spontaneous time
translation symmetry breaking provided the potential function is negative.  
}
\end{abstract}

\smallskip


\smallskip


\section{Introduction}
Recently Shapere and Wilczek \cite{SW1,SW2} have shown that for
certain special Lagrangian systems the time translation symmetry
can be spontaneously broken in the lowest energy  or ground state.
This has revived  interest in the study of  systems with
non-standard and/or non-convex Lagrangians especially with regard
to spontaneous breaking of time translation symmetry.
A direct consequence of the spontaneously broken time translation symmetry
in the ground states is the multivaluedness of the
Hamiltonian.

A common feature shared by all  the  models considered by Shapere and
Wilczek \cite{SW1,SW2} is that the energy function (Hamiltonian) or
Lagrangian systems  become multivalued in terms
of the canonical phase space variables.
 Recently it has become clear that, for special kinds of mechanical
systems, there are choices of Hamiltonian structures in which certain fundamental aspects
of classical canonical Hamiltonian mechanics are changed. It has been
explored in \cite{ZXY1,ZXY,Zhao}, one can change the phase space variables
which makes the Hamiltonian and symplectic structures on the phase space simultaneously
well defined at the price of introducing a non-canonical symplectic structure.
Curtright and Zachos \cite{Curtright}
displayed some simple unified Lagrangian
prototype systems which, by virtue of non-convexity in their velocity dependence, branch into
double-valued (but still self-adjoint) Hamiltonians.

\smallskip

It is noteworthy that for systems possessing multiple Hamiltonian descriptions, there have been
discussions in the literature as to find the proper choice of Hamiltonian functions.
Furthermore an analysis of such
models has even led to speculations about the possibility of
perpetual motion. Shapere and Wilczek papers triggered a new interest on the
systems with branched Hamiltonians.

\smallskip

The issue of time independent classical dynamical systems
exhibiting motion in their lowest energy states has been
instrumental in the introduction of  a time analogue of spatial
order as in a crystalline substance \cite{SW1} (the so called
\textit{time crystals}) and its spontaneous breaking. It is
therefore natural to investigate the issue of time translation
breaking from the perspective of second-order differential
equations within the general framework of Lagrangian/Hamiltonian
mechanics \cite{ZXY1,ZXY}.

\smallskip

{\bf Motivation and result :}
 The motivation for the present work arose originally from  Shapere and Wilczek's observation that the Lagrangians of some mechanical systems display spontaneous time translation symmetry breaking properties in their lowest energy state, and  the
Hamiltonian descriptions of certain singular models involving  multi-valuedness and  branching point singularities. In a previous article we  obtained the Chiellini integrability criterion for the Li\'enard equation
by using Jacobi's last multiplier \cite{CGDCDS} and derived the bi-Hamiltonian structure of
those equations of the Li\'enard type satisfying this particular criterion. Moreover we also constructed certain non-natural Lagrangians and Hamiltonians for the Li\'enard equation using Jacobi's last multiplier; consequently it is only natural that we investigate the possible existence of time translation symmetry breaking of the ground state for such systems.
The first case we deal with is that of  a second-order ordinary
differential equation (ODE) of the usual Li\'{e}nard type viz 
 \be\label{L1}
\ddot{x}+f(x) \dot{x}+g(x)=0,\ee for which we present specific cases of a double valued Hamiltonian and its branches. This is followed up with a quadratic
version (the Li\'{e}nard-II equation)\cite{Sabatini}, namely
 \be\label{L2}
\ddot{x}+f(x) \dot{x}^2+g(x)=0.\ee The latter  naturally
emerges from Newton's second law when dealing with a system
characterized by a variable mass (depending on the position
coordinate) and also frequently arises in the context of isochronous systems \cite{CG,GC,CG1}. By a suitable modification of the Hamiltonian of this equation we obtain the locus of the curve of the singular points for which the energy is less than the minimum value indicating the spontaneous breaking of time translation symmetry. 

\smallskip

This paper is organized as follows. We present the branched Hamiltonian description of the Li\'enard
equation in Section 2. We also illustrate the double valuedness of the Hamiltonian
description.  Section 3 is devoted to the hamiltonization of an equation 
of Lienard type with a quadratic dependence on the velocity, dubbed as 
Li\'enard II equation. 
We demonstrate how the time translation symmetry spontaneously broken for Li\'enard II system in Section 4.

\section{The Li\'{e}nard-I equation and branched Hamiltonians}
There exists an extensive literature on the Li\'{e}nard-I equation
( for example, \cite{Perko,Chicone}) and in this section our attempt is to incorporate the
Li\'{e}nard-I equation \be\label{k1}\ddot{x}+f(x)\dot{x}+g(x)=0
,\ee into the branched Hamiltonian framework. 
It has been shown in \cite{CG,GC} how a system  of the Li\'{e}nard
type as given by (\ref{L2}) can be embedded into the Hamiltonian
formalism. We briefly recapitulate the procedure below. Given a
second-order ordinary differential equation (ODE)
\bel{2.1}\ddot{x}=F(x,\dot{x})\ee we define the Jacobi last
multiplier $M$ as a solution of the following ODE
\bel{2.2}\frac{d\log M}{dt}+\frac{\partial
F(x,\dot{x})}{\partial\dot{x}}=0.\ee Assuming (\ref{2.1}) to be
derivable from the Euler-Lagrange equation one can show that the
JLM is related to the Lagrangian by the following equation
\bel{2.3}M=\frac{\partial^2L}{\partial \dot{x}^2}.\ee
From (\ref{2.2}) a formal
solution of the Jacobi last multiplier for (\ref{k1}) may be
written as \be\label{k3.8} M(t,x)=\exp\left(\int f(x)
dt\right):=u^{1/\ell}, \ee where  $u$ is a new nonlocal variable
and $\ell$ is a parameter whose value is fixed by the
 following lemma once $f$ and $g$ are given.
 \begin{lem}
The Li\'{e}nard equation (\ref{k1}) can be written
  as the following system
  $$\dot{u}=\ell uf(x),\;\;\;\dot{x}=u+W(x)$$ where
  $W=g/f\ell$ with the parameter $\ell$ being
  determined by the following condition
  \be\label{cond1}\frac{d}{dx}\left(\frac{g}{f}\right)
  =-\ell(\ell+1)f(x).\ee
\end{lem}

{\bf Proof}: From (\ref{k3.8}) we have $\log{u}=\ell\int f(x) dt$,
which implies $\dot{u}=\ell u f(x)$. Setting $\dot{x}=u+W(x)$ we
find by differentiating with respect to $t$
$$\ddot{x}=\dot{u}+W^\prime(x)\dot{x}.$$ Inserting the expression
for $\dot{u}$ from the previous equation and after eliminating $u$
we find that
$$\ddot{x}=\ell f(x)(\dot{x}-W)+W^\prime(x)\dot{x}.$$
Comparison with (\ref{k1}) then shows $W^\prime(x)=-(\ell+1)f(x)$
and $W(x)= g/\ell f$. Consistency now requires that
$$\frac{d}{dx}\left(\frac{g}{f}\right)
  =-\ell(\ell+1)f(x),$$
which represents actually the Cheillini integrability condition
for (\ref{k1}) ( see \cite{CGDCDS}, for Cheillini integrability condition in the context of Li\'enard equation).

\smallskip

Since the transformation is nonlocal so a mapping to the
$(x,u)$-plane is not possible and therefore one cannot really
analyse the problem in the local manner of point transformations.

\smallskip

 However,  from (\ref{2.3}) and (\ref{k3.8}) we have
$$\frac{\partial^2L}{\partial
\dot{x}^2}=\left(\dot{x}-\frac{1}{\ell}\frac{g}{f}\right)^{1/\ell},$$
 and it may be shown that (\ref{k1}) can be derived from the
 following Lagrangian
\be\label{Lie1Lag}
 L=\frac{\ell^2}{(\ell+1)(2\ell+1)}\left(\dot{x}-\frac{1}{\ell}\frac{g}{f}\right)^{{(2\ell+1)}/{\ell}},\ee
 provided the functions $f$ and $g$ satisfy the Cheillini
 integrability condition (\ref{cond1}).

\subsection{A class of double-valued Hamiltonians}

 Before proceeding to a determination of the Hamiltonian for
 (\ref{k1}) from the above Lagrangian  we note that
 the curvature $\partial^2L/\partial
{\dot{x}^2}$ changes sign at the points where
$\dot{x}={g}/{f\ell}$
 provided $\ell$ is an odd integer or $1/\ell$ is an odd integer. The conjugate
 momentum is given as usual by
 $$p=\frac{\ell}{\ell+1}\left(\dot{x}-\frac{1}{\ell}\frac{g}{f}\right)^{(\ell+1)/\ell}.$$
 The inversion of this relation to determine $\dot{x}$ as a function of $p$ and $x$ presents us with difficulty and is the source of the double valuedness of the resulting Hamiltonian. 
Formally the Hamiltonian is
$$ H =p^{2\ell +1/\ell+1} K(\ell)-(g/\ell f)p$$ where
$K(\ell)$ is just a scaling factor.

\smallskip

By enlarging the phase space and making
use of Dirac's theory on constrained Hamiltonian systems Zhao {\it et al} \cite{ZXY} 
presented the Hamiltonian description
and formulated a method to avoid the multivaluedness and the brunching point singularities.

We consider the following example to illustrate our point.\\

 \noindent
 \textit{Example}\\
  $\ddot{x}+x\dot{x}+x-x^3=0$\\
 Here $f(x)=x$ and $g(x)=x-x^3$. One can easily verify that the Cheillini condition is satisfied with $\ell=1$ and $-2$. For $\ell=1$ we obtain $p=(\dot{x}-1+x^2)^2/2$. A plot of the variation of the conjugate momentum with $x$ and $\dot{x}=y$ is shown below in Fig. 1 .
On the other hand upon inversion we have $\dot{x}=1-x^2\pm\sqrt{2p}$ and a plot of the variation of $\dot{x}$ with $x$ and $p$ is depicted in Fig. 1. It is observed that $\dot{x}=1-x^2\pm\sqrt{2p}$ whence the Hamiltonian is double valued with the branches:
 $$H_{\pm}=p(1-x^2\pm\frac{2}{3}\sqrt{2p})$$
\begin{figure}[ht!]
 \centering
 \includegraphics[width=100mm]{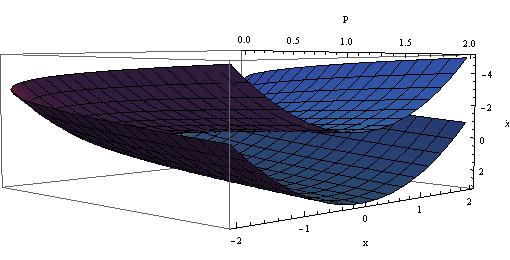}
 \caption{3D plot showing the variation  $\dot{x}=1-x^2\pm\sqrt{2p}$  when $\ell=1$, the lower (upper) one is the negative 
ne, both meet at $p=0$}
\label{fig1}\end{figure}
The variation of the Hamiltonians are depicted below in Fig 2.
\begin{figure}[ht!]
\centering
\includegraphics[width=100mm]{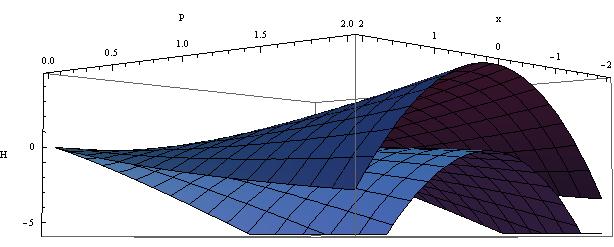}
 \caption{3D plot showing the variation  of the Hamiltonian $H_{\pm}$  when $\ell=1$}
\label{fig2}\end{figure}

 However, when $\ell=-2$ then $p=2\sqrt{\dot{x}+\frac{1}{2}(1-x^2)}$ leading to $\dot{x}=p^2/4-(1-x^2)/2$ and leads to the Hamiltonian $H=p^3/12-p(1-x^2)/2$, i.e., we have a single valued Hamiltonian.
\begin{figure}[ht!]
 \centering
\includegraphics[width=100mm]{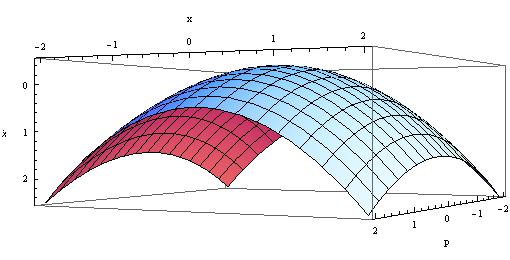}
\caption{3D plot showing the variation $\dot{x} = p^2/4 - (1-x^2)/2$ when $\ell=-2$}
\label{fig3}\end{figure}

\smallskip

We illustrate the variation of velocity and Hamiltonian when $l = -2$ in figure 3 and 4 diagrams respectively.

\smallskip

\begin{figure}[ht!]
 \centering
 \includegraphics[width=100mm]{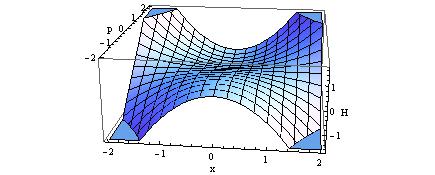}
\caption{3D plot showing the variation of Hamiltonian when $\ell=-2$}
\label{fig4}\end{figure}

\section{Hamiltonian aspects of Li\'{e}nard-II equation}

 For the
equation \bel{2.4}\ddot{x}+f(x)\dot{x}^2 +g(x)=0,\ee one can show
that a solution of the JLM is given by
\bel{2.5}M(x)=e^{2F(x)},\;\;\;F(x):=\int^x f(s) ds.\ee Furthermore
it follows  from (\ref{2.3}) that its Lagrangian is \bel{2.6}
L(x,\dot{x})=\frac{1}{2}e^{2F(x)}\dot{x}^2-V(x),\ee where the
potential term \bel{2.7} V(x)=\int^xe^{2F(s)}g(s) ds.\ee Clearly
the conjugate momentum \bel{2.8} p:=\frac{\partial
L}{\partial\dot{x}}=\dot{x}e^{2F(x)}\;\;\mbox{implies}\;\;\dot{x}=pe^{-2F(x)},\ee
so that the final expression for the Hamiltonian is
\bel{2.9}H=\frac{p^2}{2M(x)}+\int^x M(s) g(s) ds,\ee where
$p=M(x)\dot{x}$ and $M(x)=\exp(2F(x))$ with $F(x)=\int^x f(s)ds$.
The canonical variables are $x$ and $p$ and they satisfy the standard
Poisson brackets  $\{x, p\}=1$. In terms of the canonical Poisson brackets the equations of motion appear as
 $$ \dot{x}=\{x, H\}=\frac{p}{M(x)}, \;\;\;\dot{p}=\{p, H\}=\frac{M^\prime(x)}{2M(x)}p^2-M(x)g(x)$$
 from which we can recover (\ref{2.4}) upon elimination of the conjugate momentum $p$.
 Here we have purposely written the Hamiltonian $H$ in terms of
the last multiplier $M(x)$ to highlight the latter's role as a position
dependent mass term. From (\ref{2.9}) it is  natural  that the potential
$V(x)$ be identified with \be V(x)=\int^x M(s) g(s) ds.\ee As for
the existence of a minima of $H$, considered as a function of $x$
and $p$, it is necessary that \be \frac{\partial H}{\partial
x}=0\;\;\;\mbox{and}\;\;\;\frac{\partial H}{\partial p}=0\ee whose
solutions then define the stationary points. The former yields
$$-p^2\frac{M^\prime(x)}{2M^2(x)}+M(x)g(x)=0$$ while the latter
implies $p/M(x)=0$. Therefore the stationary points are
characterized by $p=0$ and the value(s) of $x$ for which $g(x)=0$.
If $x=x^\star$ denotes a root of $g(x)=0$ then $(x^\star, p=0)$ is
a stationary point (s.p). For the s.p to be a minimum one
requires that the principal minors of
$$\Delta=\left| \begin{array}{ccc}H_{xx} & H_{xp}\\H_{px} &
H_{pp}\end{array}\right|_{s.p}$$ be positive definite, i.e.,
$$ g^\prime(x^\star)>0\;\;\;\mbox{ and }\;\;\;M(x^\star)g^\prime(x^\star)>0,$$
and consistency therefore requires  $M(x^\star)>0$. Note that
$M(x)$, which may be thought of as some kind of 'effective mass'
such as within a spatial crystal, may be negative for $x\ne
x^{\star}$. Clearly the fact that $p=0$ in the minimum energy
state (ground state) of the system precludes the possibility of
any motion.

\section{A modified Hamiltonian  and spontaneously broken time
translation symmetry}

Consider a one-dimensional generalized Hamiltonian system $\widetilde{H} = {\cal F}(H)$
with Hamiltonian vector field given in terms of the canonical form
$$
{\Bbb X}_{\widetilde{H}} = \frac{\partial {\widetilde{H}}}{\partial p}\frac{\partial}{\partial x} -
\frac{\partial {\widetilde{H}}}{\partial x}\frac{\partial}{\partial p}, \qquad \{G,{\widetilde{H}}\} = \dot{G}.
$$
In the symplectic coordinates $(x,p)$ this is equivalent to canonical Hamiltonian equations
$$ \dot{x} = {\cal F}(H)^{\prime}\{x,H\}, \qquad \dot{p} = {\cal F}(H)^{\prime}\{p,H\}, \qquad \hbox{ where }\,\,\,\,   {\cal F}(H)^{\prime} > 0. $$
It may be easily verified that the above set of Hamiltonian equations may be obtained from the modified symplectic form
$\omega =  {\cal F}(H)^{\prime} dx \wedge dp$. Moreover this change of Hamiltonian structure
will not change the partition function, hence all thermodynamic quantities will remain unchanged.

\noindent
 Let us  consider  a new Hamiltonian \cite{ZXY} defined by
\be\label{H2} \widetilde{H}=\left(\frac{p^2}{2M(x)}+\int^x M(s)
g(s) ds\right)^2+E_0=H^2+E_0,\ee where $E_0$ is an arbitrary constant. As the New Hamiltonian is anticipated to generate a dynamics which is distinct from that of $H$, let us also introduce the following Poisson structure
 $\{x, p\}=\xi(x,p)$ so that the equations of motion which follow from
\be\label{E1}
\dot{x}=\{x,\widetilde{H}\},\;\;\;\dot{p}=\{p,\widetilde{H}\}\ee
 give \be\dot{x}=2\xi H\frac{p}{M(x)}\ee \be\dot{p}= -2\xi
H\left(-\frac{M^\prime(x)}{2M^2(x)}p^2+M(x) g(x)\right).\ee At this point we need to make a clear distinction regarding the two Poisson structures we have introduced. It will be noticed that if one assumes $\{x, p\}=\xi(x, p)=\frac{1}{2H(x,p)} $ then we get
back the original Li\'{e}nard-II equation (\ref{2.4}), if however we persist
with $\xi=1$, i.e., assume $x$ and $p$ are canonical then the
equation of motion resulting from the Hamiltonian $\widetilde{H}$
is of the form \be\label{E3}\ddot{x}+2H(f(x)\dot{x}^2+g(x))=0.\ee
Although (\ref{E3}) appears to be different from (\ref{2.4}) it is interesting to note that (\ref{E3})  can be mapped to the original set of
 Hamiltonian equations by using a (nonlocal) Sundman transformation \cite{Sundman} through a transformation of the independent temporal variable $t$ to a
new independent variable $s$ given by $ds = 2H dt$, whence we obtain
\be {x}^{\prime}= \frac{p}{M(x)}, \qquad {p}^{\prime} = -\left(-\frac{M^\prime(x)}{2M^2(x)}p^2+M(x) g(x)\right),\ee
where $^\prime = \frac{d}{ds}$. In fact such transformations were used by Sundman while attempting to solve the restricted  three body problem.
\smallskip

As for the stationary points of the  Hamiltonian $\widetilde{H}$, these follow
from the solutions  of $\partial\widetilde{H}/\partial x=0$ and
$\partial\widetilde{H}/\partial p=0$. The latter yields either
$p=0$ or $H=0$. If $p=0$ then the former condition gives either
$H=0$ or $g(x)=0$, i.e $x=x^\star$. The pair $(x^\star, p=0)$
leads by the previous analysis to the case \be\label{E4}
\widetilde{H}_{min}=\left(\int^{x^\star} M(s) g(s)
ds\right)^2+E_0.\ee
From the above equation it is clear that the local minimum of $\widetilde{H}$ is in general greater than
the constant $E_0$ because the potential $V(x^\star)$ is not required to vanish at $x=x^\star$. As the stationary point corresponds to $p=0$ the time translation symmetry is not broken and we have the same situation as previously discussed in section 2.

However one also has now the possibility
wherein $H=0$ which implies that the locus of the stationary points lie on the
curve
\be\label{E3.7a}\frac{p^2}{2M(x)}+\int^x M(s) g(s) ds=0.\ee This condition
obviously implies that $\widetilde{H}$ has a minima with
$\widetilde{H}_{min}=E_0$ which is  less than that given by
(\ref{E4}). Now for real values of $p$ it is then necessary that
$$V(x)=\int^x M(s) g(s) ds<0.$$
The force $dV/dx$ is clearly not necessarily zero and motion can therefore occur in the ground state.
The existence of motion under such
circumstances is indicative of the spontaneous breaking of the
time-translation symmetry \cite{SW1}.

\bigskip

To investigate the possible nature of the motion in this scenario
let us demand that
\be\label{E3.7b}V(x)=\int^x M(s) g(s) ds=-\frac{1}{2}X(x)^2,\ee where $X(x)=\int
\sqrt{M(x)} dx$. Such a choice is consistent with the view expressed in \cite{ZXY1} that time translation symmetry may be present in almost all Newtonian mechanical systems with a conservative potential provided the potential can be shifted to acquire a negative value. Furthermore such symmetry breaking occurs in a non-standard Hamiltonian description where the new Hamiltonian is the square of the canonical Hamiltonian together with Poisson brackets which are nonlinear.
Differentiating (\ref{E3.7b})we get
$$M(x)g(x)=-X(x)X^\prime(x) \;\;\;\mbox{with}\;\;X^\prime(x)=\sqrt{M(x)}=e^{F(x)}$$
so that $e^{F(x)}g(x)=-X(x)$ which after another differentiation with respect to $x$
 leads to the
condition  \be\label{inviso} g^\prime(x)+f(x) g(x)=-1,\ee in view of the fact that $f(x)=M^\prime(x)/2M(x)$. Notice
that this basically represents motion in a inverted oscillator
potential and it is therefore not surprising that  the last
condition on the functions $f$ and $g$ is just the `inverted
isochronicity' condition \cite{Sabatini}. The notion of an inverted oscillator also appears in the context of
de-Sitter gravity. To arrive at concrete
models for the function $f$ in this case, we note that one may
solve (\ref{inviso}) for $f$ to get \be
f=-\frac{1+g^\prime}{g},\;\;\;\mbox{which then implies
}\;M(x)=\frac{1}{g^2(x)}\exp\left(-2\int\frac{dx}{g}\right).\ee from (\ref{E3.7b}) it follows that
\be X(x)=\int
\frac{1}{g(x)}\exp\left(-\int\frac{dx}{g}\right)dx\ee

The points of minima therefore lie on the curve
$$p=\pm \sqrt{M(x)}X(x)=\pm
\frac{1}{g(x)}\exp\left(-\int\frac{dx}{g}\right)\int
\frac{1}{g(x)}\exp\left(-\int\frac{dx}{g}\right)dx.$$
We end this section with a couple of examples:\\
\textit{Example 1}\\
Let $g(x)=x$  then we have
$$M(x)=\frac{1}{x^4}, \;\;X(x)=-x,\;\; \mbox{and}\;\; p=\mp \frac{1}{x^3},$$ the
 singular nature of $M(x)$ at $x=0$ forces us to confine ourselves
 to the half line. It is evident that the particle can at any instant of time have only one of the two
 possible values for the momentum. The particular choice of any one of these two possible
 values therefore breaks the time translation symmetry.\\
 \textit{Example 2}\\
  If $g(x)=1/x$ then we obtain
 $$M(x)=x^2e^{-\frac{x^2}{2}},\;\;X(x)=-e^{-\frac{x^2}{2}}\;\;\;\mbox{and}\;\;\;p=\mp
 x e^{-x^2}.$$

\section{Conclusion}

We have shown that nonlinear ODEs of the Li\'enard
type it is easy to recast
them into the Lagrange/Hamilton formalism and the basic results of
Shapere-Wilczek are apparently applicable to such a differential
system. In particular, we have studied the branched Hamiiltonian and multivaluednes of momentum of this
equation. Our analysis is based on $\widetilde{H} = H^2 + E_0$.
Actually when we consider such kind of generalized Hamiltonian the number of critical points is changed
drastically, and most of the
critical points of the generalized Hamiltonian are not the images of the critical point of the original Hamiltonian.
Careful readers might have noticed
that this (quadratic) Hamiltonian connected to (exotic) Lagrangian via Legendre transformation
injected the multivaluedness of the momentum.

\smallskip

Shapere and Wilczek
found that the  direct consequence of this multivaluedness is that
the time translation symmetry is spontaneously broken in the ground states.
The phenomenon of spontaneous symmetry breaking  was hitherto
mostly restricted to the quantum domain. The most outstanding
example being that of the Higgs boson besides superconductors,
ferromagnets and liquid crystals. The fact that such a phenomenon
may also occur in the classical regime is tantalizing at least
from the theoretical point of view if nothing else. The
introduction of the associated  concept of time crystals by
Shapere and Wilczek is not without controversy especially
regarding their experimental realization. While the examples
considered by them as also by  L. Zhao \textit{et al} are drawn
from classical mechanics and field theory our motivation in this
note is to extend this notion to nonlinear ordinary differential
equations. We have shown that nonlinear ODE of the Li\'enard
type it is easy to recast
them into the Lagrange/Hamilton formalism and the basic results of
Shapere-Wilczek are apparently applicable to such a differential
system.

\section*{ Acknowledgement}
We are extremely grateful to Professor Liu Zhao for his
his interest and valuable suggestions. We would also like to thank Ankan Pandey for the diagrams.
Finally, we would like to thank the anonymous referee for carefully reading our manuscript 
and for giving constructive comments.

\end{document}